\documentclass{PoS}

\usepackage{amsthm,amsmath,amssymb}
\usepackage{rotating}

\newcommand{\C}{\mathbb C}

\newcommand{\eg}{e.g.,\ }
\newcommand{\ie}{i.e.,\ }

\newcommand\Mvec{\,\mbox{\bf M}}

\newcommand{\HarmonicSumsP}{\texttt{HarmonicSums}}

\theoremstyle{plain}
\newtheorem{theorem}{Theorem}

\theoremstyle{definition}

\newtheorem{example}[theorem]{Example}

\theoremstyle{remark}

\newcounter{mmacnt}
\def\restartmma{\setcounter{mmacnt}{0}}
\restartmma \catcode`|=\active
\def|#1|{\mathrm{#1}}
\catcode`|=12

\allowdisplaybreaks[4]

\title{Computing the Inverse Mellin Transform of Holonomic Sequences using Kovacic's Algorithm}

\ShortTitle{Inverse Mellin Transform of Holonomic Sequences}

\author{\speaker{Jakob Ablinger}\thanks{This work has been supported in part by the Austrian Science Fund (FWF) grant SFB F50 (F5009-N15).}\\
        Research Institute for Symbolic Computation (RISC)\\
	Johannes Kepler University Linz, Altenberger Stra\ss e 69, A-4040 Linz, Austria \\
        E-mail: \email{jablinge@risc.jku.at}}

\abstract{We describe how the extension of a solver for linear differential equations by Kovacic's algorithm helps to improve a method to compute the inverse Mellin transform of holonomic sequences. The method is implemented in the computer algebra package \HarmonicSumsP.}

\FullConference{13th International Symposium on Radiative Corrections\\
		24-29 September, 2017\\
		St. Gilgen, Austria}

\begin{document}

\section{Introduction}
\label{sec:1}
\noindent
There have been several methods proposed to compute the inverse Mellin transform of special sequences, for instance in~\cite{Remiddi2000} an algorithm (using rewrite rules) to compute the inverse Mellin transform of harmonic sums was stated. This algorithm was
extended in~\cite{Ablinger:2013hcp} to generalized harmonic sums such as S-sums and cyclotomic sums. A different approach to compute inverse Mellin transforms of binomial sums was described in~\cite{Ablinger:2014}.
In \cite{InvMellin} a method to compute the inverse Mellin transform of general holonomic sequences was described. That method uses holonomic closure properties and was implemented in the computer algebra package \HarmonicSumsP~\cite{HarmonicSums,Ablinger:2013hcp,Ablinger:2011te,Ablinger:2013cf,Ablinger:2010kw}. In the frame of the method a linear differential equation has to be solved.
So far the differential equations solver of \HarmonicSumsP\ was only able to find d'Alembertian solutions \cite{Abramov:94}. Recently the solver was generalized and therefore more general inverse Mellin transforms can be computed. 

In the following we repeat important definitions and properties (compare \cite{InvMellin,Ablinger:2014,KauersPaule:2011}).
Let $\mathbb K$ be a field of characteristic~0. A function $f=f(x)$ is called \textit{holonomic} (or \textit{D-finite}) if there exist 
polynomials $p_d(x),p_{d-1}(x),\ldots,p_0(x)\in \mathbb K[x]$  (not all $p_i$
being $0$) such that the following holonomic differential equation holds:
\begin{equation}
 p_d(x)f^{(d)}(x)+\cdots+p_1(x)f'(x)+p_0(x)f(x)=0.
\end{equation}
We emphasize that the class of holonomic functions is rather large due to its
closure properties. Namely, if we are given two such differential
equations that contain holonomic functions $f(x)$ and $g(x)$ as solutions, one
can compute holonomic differential equations that contain $f(x)+g(x)$,
$f(x)g(x)$ or $\int_0^x f(y)dy$ as solutions. In other words any composition
of these operations over known holonomic functions $f(x)$ and $g(x)$ is again a
holonomic function $h(x)$. In particular, if for the inner building blocks
$f(x)$ and $g(x)$ the holonomic differential equations are given, also the holonomic
differential equation of $h(x)$ can be computed.\\
Of special importance is the connection to recurrence
relations. A sequence $(f_n)_{n\geq0}$ with $f_n\in\mathbb K$ is called
holonomic (or \textit{P-finite}) if there exist polynomials 
$p_d(n),p_{d-1}(n),\ldots,p_0(n)\in \mathbb K[n]$ (not all $p_i$ being $0$) such
that a holonomic recurrence
\begin{equation}
 p_d(n)f_{n+d}+\cdots+p_1(n)f_{n+1}+p_0(n)f_n=0
\end{equation}
holds for all $n\in\mathbb N$ (from a certain point on).
In the following we utilize the fact that holonomic functions are
precisely the generating functions of holonomic sequences: 
if $f(x)$ is holonomic, then the coefficients 
$f_n$ of the formal power series expansion 
$$f(x) = \sum\limits_{n=0}^{\infty} f_n x^n$$
form a holonomic sequence. Conversely, for a given holonomic sequence
$(f_n)_{n\geq0}$, the function defined by the above sum (\ie its 
generating function) is holonomic (this is true in the sense of formal power series, even if the sum has a zero radius of 
convergence). Note that given a holonomic differential equation for a holonomic function $f(x)$ it is straightforward to 
construct a 
holonomic recurrence for the coefficients of its power series expansion. For a
recent overview of this holonomic machinery and further literature we
refer to~\cite{KauersPaule:2011}.
An additional property of holonomic functions was given for example in \cite{InvMellin} and \cite{Ablinger:2014}:
if the Mellin transform of a holonomic function
\begin{equation}\label{Equ:MellRep}
\Mvec[f(x)](n):=\int_0^1x^nf(x)dx=F(n).
\end{equation}
of a holonomic function is defined \ie the integral $\int_0^1x^nf(x)dx$ exists, then it is 
a holonomic sequence.
Conversely, if the Mellin transform $\Mvec[f(x)](n)$ of a function $f(x)$ is holonomic, then also the function $f(x)$ is holonomic.
In this article we will report an extension of \HarmonicSumsP\ based on Kovacic's algorithm \cite{Kovacic} that supports the user to calculate the inverse Mellin transform in terms of iterated integrals that exceed the class of d'Alembertian solutions.

The paper is organized as follows. In Section \ref{sec:2} we revisit the method to compute the inverse Mellin transform of holonomic functions from \cite{InvMellin}, while in Section \ref{sec:3} we explain the generalization of the method and in Section \ref{sec:4} we give some examples.

\section{The Inverse Mellin Transform of Holonomic Sequences}
\label{sec:2}
\noindent
In the following, we deal with the following problem:\\
\textbf{Given} a holonomic sequence $F(n)$.\\
\textbf{Find,} whenever possible, a holonomic function $f(x)$ such that for all $n\in\mathbb N$ (from a certain point on) we have
\begin{equation*}
\Mvec[f(x)](n)=F(n).
\end{equation*}
In \cite{InvMellin} a procedure was described to compute a differential equation for~$f(x)$ given a holonomic recurrence for $\Mvec[f(x)](n).$
Given this procedure the following method to compute the inverse Mellin transform of holonomic sequences was proposed in \cite{InvMellin}:
\begin{enumerate}
 \item Compute a holonomic recurrence for $\Mvec[f(x)](n).$\label{holorec}
 \item Use the method mentioned above to compute a holonomic differential equation for $f(x).$
 \item Compute a linear independent set of solutions of the holonomic differential\\
       equation (using~\HarmonicSumsP).\label{DiffSol}
 \item Compute initial values for $\Mvec[f(x)](n).$
 \item Combine the initial values and the solutions to get a closed form
representation for~$f(x).$
\end{enumerate}
\noindent In our applications we usually apply this method on expressions in terms of nested sums, however as long as there is a method to compute the holonomic recurrence for a 
given expression (\ie item~\ref{holorec} can be performed) this proposed method can be used. Another possible input would be a holonomic recurrence together with sufficient initial values.
Note that until recently $\HarmonicSumsP$ could find all solutions of holonomic differential equations that can be expressed in
terms of iterated integrals over hyperexponential alphabets~\cite{Ablinger:2014,Bronstein,Singer:99,Petkov:92}; these solutions
are called d'Alembertian solutions \cite{Abramov:94}. Hence as long as
such solutions were sufficient to solve the differential equation in item~\ref{DiffSol} we succeeded to compute~$f(x).$ In case d'Alembertian solutions
do not suffice to solve the differential equation in item~\ref{DiffSol}, we have to extend the solver for differential equations.

\section{Beyond d'Alembertian solutions of linear differential equations}
\label{sec:3}
\noindent
Until recently only d'Alembertian solutions of linear differential equations could be found in using \HarmonicSumsP\ (compare \cite{InvMellin}), but in order to treat more general problems the differential equation solver had to be extended. 
In \cite{Kovacic} an algorithm to solve second order linear homogeneous differential equations is described. We will refer to this algorithm as Kovacic's algorithm.

\noindent Consider the \textit{holonomic} differential equation ($p_i(x)\in\C[x]$)
\begin{equation}\label{KovacicDE}
    p_2(x)f''(x)+p_1(x)f'(x)+p_0(x)f(x)=0.
\end{equation}
Kovacic's algorithm decides whether (\ref{KovacicDE})
\begin{itemize}
 \item has a solution of the form $e^{\int\omega}$ where $\omega\in\C(x)$;
 
 \item has a solution of the form $e^{\int\omega}$ where $\omega$ is algebraic over $\C(x)$ of degree 2 and the previous case does not hold;

 \item all solutions are algebraic over $\C(x)$  and the previous cases do not hold;

 \item has no such solutions;
\end{itemize}
and finds the solutions if they exist. Note that the solutions Kovacic's algorithm can find are called Liouvillian solutions \cite{Singer:99}. In case Kovacic's algorithm finds a solution, it is straightforward to compute a second solution which will again be Liouvillian.
This algorithm was implemented in~\HarmonicSumsP.

\begin{example}
Consider the following differential equation:
\begin{eqnarray*}
  &&\biggl(4 x \left(40-891 x+1701 x^2\right)+9 x^2 \left(32-376 x+459 x^2\right) D_x\nonumber+18 x^3 \left(4-31 x+27 x^2\right) D_x^2\biggr) f(x)=0,
\end{eqnarray*}
  with the implementation of Kovacic's algorithm in \HarmonicSumsP\ we find the following two solutions:
\begin{eqnarray*}
   f_1(x)&=&-\frac{\sqrt{10-6 \sqrt{1-x}-x} \sqrt[6]{2+2 \sqrt{1-x}-x}}{\sqrt{(1-x)^3} \sqrt[3]{x^5} (-4+27 x)},\\
   f_2(x)&=&-\frac{\sqrt{10+6 \sqrt{1-x}-x} \sqrt[6]{2-2 \sqrt{1-x}-x}}{\sqrt{(1-x)^3} \sqrt[3]{x^5} (-4+27 x)}.
\end{eqnarray*}
\end{example}

\subsection{Composing solutions}
\noindent Suppose we are given the linear differential equation ($q_i,p_i\in\C[x];d>0$)
\begin{eqnarray}
    \left(q_{d}(x)D_x^{d}+\cdots+q_0(x)\right)f(x)&=&0,\label{composesoldeq1}
\end{eqnarray}
which factorizes linearly into $d$ first-order factors. Then this yields $d$ linearly independent solutions of the form
$$
f_1(x),f_1(x)\int{\frac{f_2(x)}{f_1(x)}}dx,f_1(x)\int{\frac{f_2(x)}{f_1(x)}\int{\frac{f_3(x)}{f_2(x)}}dx}dx ,\ldots, f_1(x)\int{\frac{f_2(x)}{f_1(x)}\cdots \int{\frac{f_d(x)}{f_{d-1}(x)}} dx}\cdots dx,
$$
where the $f_i$ are hyperexponential functions (\ie $\frac{D_x f_i(x)}{f_i(x)}\in \mathbb K(x)^*$). These solutions are also called d'Alembertian solutions of (\ref{composesoldeq1}), compare \cite{Petkov:92,Abramov:94}.

Now suppose that a given differential equation does not factorize linearly, but contains in between second-order factors, which can be solved, \eg by Kovacic's algorithm.
Let the following differential equation correspond to a second order factor: 
\begin{eqnarray}
    \left(p_2(x)D_x^2+p_1(x)D_x+p_0(x)\right)f(x)&=&0,\label{composesoldeq2}
\end{eqnarray}
then we can compose the solutions of the first order and second order factors as follows.
Let $s(x)$ be solution of (\ref{composesoldeq1}) and let $g_1(x)$ and $g_2(x)$ be solutions of (\ref{composesoldeq2}). Then 
  $$s(x),\ s(x)\int{\frac{g_1(x)}{s(x)}dx}\text{ and }s(x)\int{\frac{g_2(x)}{s(x)}dx}$$ 
  are solutions of
  $$\left(p_2(x)D_x^2+p_1(x)D_x+p_0(x)\right)\left(q_{d}(x)D_x^{d}+\cdots+q_0(x)\right)f(x)=0.$$
In addition, if we define $w(x):=p_2(x)(g_1'(x)g_2(x)-g_1(x)g_2'(x))$ then 
  $$g_1(x),\ g_2(x)\text{ and }g_1(x)\int s(x) w(x) g_2(x)dx-g_2(x)\int s(x) w(x) g_1(x)dx$$ 
are solutions of
  $$\left(q_{d}(x)D_x^{d}+\cdots+q_0(x)\right)\left(p_2(x)D_x^2+p_1(x)D_x+p_0(x)\right)f(x)=0.$$

\section{Examples}
\label{sec:4}
\begin{example}We want to compute the inverse Mellin transform of
 $$f_n:=\left(\frac{4}{27}\right)^n\binom{3n}{n}.$$
 We find that
\begin{eqnarray*}
-2 (3 n+1) (3 n+2) f_{n}+ 9 (n+1)(2 n + 1) f_{n+1}=0,
\end{eqnarray*}
which leads to the differential equation
\begin{eqnarray*}
(27 x - 4) f(x)+9 x (7 x -4) f'(x)+18 x^2 (x -1) f''(x)=0,
\end{eqnarray*}
for which we find with the help of Kovacic's algorithm the general solution
$$s(x)=c_1\frac{\sqrt[6]{2+2 \sqrt{1-x}-x}}{\sqrt{1-x} \sqrt[3]{x^2}}+c_2 \frac{\sqrt[6]{2-2 \sqrt{1-x}-x}}{\sqrt{1-x} \sqrt[3]{x^2}},$$
for some constants $c_1$ and $c_2.$ In order to determine these constants we compute 
\begin{eqnarray*}
\int_0^1x^1s(x)dx&=&\frac{1}{9} c_1 \left(-3+\frac{8 \pi }{\sqrt{3}}+8 \log (2)\right)+\frac{1}{9} c_2 \left(3+\frac{8 \pi }{\sqrt{3}}-8 \log (2)\right),\\
\int_0^1x^2s(x)dx&=&\frac{1}{486} c_1 \left(-147+\frac{320 \pi }{\sqrt{3}}+320 \log (2)\right)+\frac{1}{486} c_2 \left(147+\frac{320 \pi }{\sqrt{3}}-320 \log (2)\right).
\end{eqnarray*}
Since $f_1=4/9$ and $f_2=80/243$ we can deduce that $c_1=c_2=\frac{\sqrt{3}}{4 \pi }$ and hence
$$
f_n=\frac{\sqrt{3}}{4 \pi }\Mvec\left[\frac{\sqrt[6]{2-2 \sqrt{1-x}-x}+\sqrt[6]{2+2 \sqrt{1-x}-x}}{\sqrt{1-x} x^{2/3}}\right](n).
$$
\end{example}

\begin{example}
 During the Computation of the inverse Mellin transform of
 $$\sum _{i=1}^n \binom{3i}{i}\frac{1}{i}$$
 we have to solve the following differential equation:
 \begin{eqnarray*}
  0&=&27 (27 x-4) f(x)+\left(4131 x^2-2160 x+16\right) f'(x)+9 x \left(351 x^2-298 x+16\right) f''(x)\\&&+18 (x-1) x^2 (27 x-4) f^{(3)}(x).
 \end{eqnarray*}
 We are able to find the general solution of that differential equation using \HarmonicSumsP:
 \begin{eqnarray*}
    c_1\frac{1}{27 x-4}
   +c_2\frac{\int_0^x \frac{1}{\left(1-\sqrt{1-\tau }\right)^{2/3} \sqrt[3]{1+\sqrt{1-\tau }} \sqrt{1-\tau }} \, d\tau }{27 x-4}
   +c_3\frac{\int_0^x \frac{1}{\sqrt[3]{1-\sqrt{1-\tau }} \left(1+\sqrt{1-\tau}\right)^{2/3} \sqrt{1-\tau }} \, d\tau }{27 x-4}.
 \end{eqnarray*}  
 Given this general solution we find:
 \begin{eqnarray*}
 \sum _{i=1}^n \binom{3i}{i}\frac{1}{i}
    &=& \left(\frac{27}{4}\right)^{n+1}\Biggl(4\int_0^1 \left(x^n-\frac{4^n}{27^n}\right)\frac{1}{27 x-4} \, dx\\
    &&-\frac{\sqrt{3}}{\pi} \int_0^1 \left(x^n-\frac{4^n}{27^n}\right) \frac{\displaystyle{\int_0^x{\frac{1}{\sqrt[3]{1-\sqrt{1-\tau }} \left(1+\sqrt{1-\tau }\right)^{2/3} \sqrt{1-\tau }}d\tau}}}{27 x-4} \, dx\\
    &&-\frac{\sqrt{3}}{\pi} \int_0^1 \left(x^n-\frac{4^n}{27^n}\right) \frac{\displaystyle{\int_0^x{\frac{1}{\sqrt[3]{1+\sqrt{1-\tau }} \left(1-\sqrt{1-\tau }\right)^{2/3}\sqrt{1-\tau }}d\tau}}}{27 x-4} \, dx\Biggr).
 \end{eqnarray*}
\end{example}
Finally, we list several examples that could be computed  using \HarmonicSumsP:
\begin{example}
 \begin{eqnarray*}
 \binom{4n}{2n}&=&16^n\frac{1}{2\sqrt{2}\pi}\int_0^1{\frac{x^n(1+\sqrt{x}+\sqrt{x-1})}{\sqrt{\sqrt{x}+\sqrt{x-1}}\sqrt{1-x}\, x^{\frac{3}4}}}\, dx,\\
 \frac{1}{n \binom{4 n}{2 n}}&=&\frac{1}{16^n \sqrt{2}}\int_0^1 \frac{x^n \left(1+\sqrt{x-1}+\sqrt{x}\right)}{\sqrt{\sqrt{x-1}+\sqrt{x}} \sqrt{1-x} x} \, dx,\\
 \frac{1}{n \binom{3 n}{n}}&=&\frac{\left(\frac{4}{27}\right)^n}{\sqrt{3}}\int_0^1 \frac{x^n \left(1+\left(\sqrt{x-1}+\sqrt{x}\right)^{2/3}\right)}{\sqrt[3]{\sqrt{x-1}+\sqrt{x}} \sqrt{1-x} x} \, dx,\\
 \sum _{i=1}^n \binom{3i}{i}&=&\frac{\sqrt{3} \left(\frac{27}{4}\right)^{n+1}}{\pi }\int_0^1 \frac{\left(x^n-\left(\frac{4}{27}\right)^n\right) \left(\left(\sqrt[6]{2-2 \sqrt{1-x}-x}+\sqrt[6]{2+2 \sqrt{1-x}-x}\right) \sqrt[3]{x}\right)}{\sqrt{1-x} (27 x-4)} \, dx,\\
 \sum _{i=1}^n \binom{4i}{2i}\frac{1}{i}&=&16^{n+1} \Biggl(\frac{\sqrt{2}}{4 \pi }\int _0^1\frac{x^n-16^{-n}}{1-16 x}\int _0^x\frac{ 1+\sqrt{y-1}+\sqrt{y}}{\sqrt{\sqrt{y-1}+\sqrt{y}} \sqrt{1-y} y^{3/4}}dydx-\int_0^1\frac{x^n-16^{-n}}{1-16 x} \, dx\Biggr)\\
 \sum _{i=1}^n \binom{3i}{i}\frac{1}{i^2}&=&\left(\frac{27}{4}\right)^{n+1}\Biggl(
  \frac{\sqrt{3}}{\pi} \int_0^1\frac{x^n-\left(\frac{4}{27}\right)^n}{27 x-4}\int_0^x\frac{1}{y} \int_0^y \frac{\sqrt[3]{1-\sqrt{1-z}}+\sqrt[3]{1+\sqrt{1-z}}}{\sqrt{1-z} z^{2/3}}\, dz \, dy \, dx,\\
  &&-4 \int_0^1 \frac{x^n-\left(\frac{4}{27}\right)^n}{27 x-4}\log (x) \, dx-4 \log \left(\frac{27}{4}\right) \int_0^1 \frac{x^n-\left(\frac{4}{27}\right)^n}{27 x-4} \, dx\Biggr).\\
 \end{eqnarray*}
\end{example}

\subsection*{Acknowledgements}
I want to thank C. Schneider, C. Raab and J. Bl\"umlein for useful discussions.


\begin{thebibliography}{99}

\bibitem{HarmonicSums}
  J.~Ablinger,
  \emph{The package HarmonicSums: Computer Algebra and Analytic aspects of Nested Sums},
  PoS LL {\bf 2014} (2014) 019
  [arXiv:1407.6180 [cs.SC]].


\bibitem{InvMellin}
  J.~Ablinger,
  \emph{Inverse Mellin Transform of Holonomic Sequences},
  PoS LL {\bf 2016} (2016) 067
  [arXiv:1606.02845 [cs.SC]].
  
\bibitem{Ablinger:2013hcp}
  J.~Ablinger,
  \emph{Computer Algebra Algorithms for Special Functions in Particle Physics},
  {\em PhD Thesis, J.~Kepler University Linz}, 2012.
  arXiv:1305.0687 [math-ph].


\bibitem{Ablinger:2014}
  J.~Ablinger, J.~Bl\"umlein, C.~G.~Raab and C.~Schneider,
  \emph{Iterated Binomial Sums and their Associated Iterated Integrals},
  J.\ Math.\ Phys.\  {\bf 55} (2014) 112301
  doi:10.1063/1.4900836
  [arXiv:1407.1822 [hep-th]].


\bibitem{Ablinger:2011te}
  J.~Ablinger, J.~Bl\"umlein and C.~Schneider,
  \emph{Harmonic Sums and Polylogarithms Generated by Cyclotomic Polynomials},
  J.\ Math.\ Phys.\  {\bf 52} (2011) 102301
  doi:10.1063/1.3629472
  [arXiv:1105.6063 [math-ph]].

  
\bibitem{Ablinger:2013cf}
  J.~Ablinger, J.~Bl\"umlein and C.~Schneider,
  \emph{Analytic and Algorithmic Aspects of Generalized Harmonic Sums and Polylogarithms},
  J.\ Math.\ Phys.\  {\bf 54} (2013) 082301
  doi:10.1063/1.4811117
  [arXiv:1302.0378 [math-ph]].
  

\bibitem{Ablinger:2010kw}
  J.~Ablinger,
  \emph{A Computer Algebra Toolbox for Harmonic Sums Related to Particle Physics},
  arXiv:1011.1176 [math-ph].

\bibitem{Abramov:94}
S.A.~Abramov and M.~Petkov{\v s}ek, 
  \textit{D'{A}lembertian solutions of linear differential and difference equations},
  in proceedings of \emph{ISSAC'94}, ACM Press, 1994.

\bibitem{Bronstein}
    M.~Bronstein,
    \emph{Linear Ordinary Differential Equations: breaking through the order 2 barrier},
    in proceedings of \emph{ISSAC'92}, ACM Press, 1992.

\bibitem{Singer:99}
  P.A. Hendriks and M.F. Singer, 
  \emph{Solving difference equations in finite terms},
  \emph{J.~Symbolic Comput.} {\bf 27}, 1999.
  
\bibitem{KauersPaule:2011}
  M.~Kauers and P.~Paule, 
  \emph{The Concrete Tetrahedron}, Text and Monographs in Symbolic Computation, Springer, Wien, 2011.
  

\bibitem{Kovacic}  
  J.J.~Kovacic,
  \emph{An algorithm for solving second order linear homogeneous differential equations},
  \emph{J.~Symbolic Comput.} {\bf 2}, 1986.

\bibitem{Petkov:92}
  M.~Petkov{\v s}ek, 
  \emph{Hypergeometric solutions of linear recurrences with polynomial coefficients},
  \emph{J.~Symbolic Comput.} {\bf 14}, 1992.
  
\bibitem{Remiddi2000}
   E.~Remiddi and J.A.M.~Vermaseren,
   \emph{Harmonic Polylogarithms}, 
   \emph{Int. J. Mod. Phys.} {\bf A15}, 2000.
   [arXiv:hep-ph/9905237v1]
  
  
\end{thebibliography}
\end{document}